\documentclass[letterpaper]{article} % DO NOT CHANGE THIS
\usepackage{aaai2026}  % DO NOT CHANGE THIS
\usepackage{times}  % DO NOT CHANGE THIS
\usepackage{helvet}  % DO NOT CHANGE THIS
\usepackage{courier}  % DO NOT CHANGE THIS
\usepackage[hyphens]{url}  % DO NOT CHANGE THIS
\usepackage{graphicx} % DO NOT CHANGE THIS
\usepackage{natbib}  % DO NOT CHANGE THIS AND DO NOT ADD ANY OPTIONS TO IT
\usepackage{caption} % DO NOT CHANGE THIS AND DO NOT ADD ANY OPTIONS TO IT
\usepackage{multirow}
\usepackage{amsmath,amsfonts}
\usepackage{booktabs}

\usepackage{newfloat}
\usepackage{listings}

\usepackage{algorithm}
\usepackage{algpseudocode}
\DeclareCaptionStyle{ruled}{labelfont=normalfont,labelsep=colon,strut=off} % DO NOT CHANGE THIS
\floatstyle{ruled}
\newfloat{listing}{tb}{lst}{}
\floatname{listing}{Listing}
\title{Disentangling Adversarial Prompts: A Semantic-Graph Defense for Robust LLM Security}

\author{
    Xiang Fang\textsuperscript{\rm 1}, Wanlong Fang\textsuperscript{\rm 2}\thanks{Corresponding Author. }
}
\affiliations{
    \textsuperscript{\rm 1}School of Software Engineering, Huazhong University of Science and Technology\\ 
    \textsuperscript{\rm 2}Nanyang Technological University, Singapore
     xfang9508@gmail.com,  wanlongfang@gmail.com
}

\begin{document}
\maketitle

% As a general rule, do not put math, special symbols or citations
% in the abstract
\begin{abstract}
Large Language Models (LLMs) are increasingly vulnerable to adversarial prompts that exploit semantic ambiguities to bypass safety mechanisms, resulting in harmful or inappropriate outputs. Such attacks, including jailbreaking and prompt injection, pose significant risks to the integrity and availability of LLMs in security-critical applications. This paper proposes the Adversarial Prompt Disentanglement (APD) framework, a novel defense mechanism that proactively identifies and neutralizes malicious components in input prompts before they are processed by the LLM. The APD framework integrates three key innovations: (1) a mutual information-based semantic decomposition method to isolate adversarial and benign prompt components, ensuring statistical independence; (2) a graph-based intent classification approach that leverages spectral analysis to detect malicious patterns in prompt semantics; and (3) a lightweight transformer-based classifier trained on real-world datasets of toxic and jailbreaking prompts, enabling efficient and accurate adversarial intent detection. Evaluated on diverse datasets containing adversarial prompts, APD demonstrates superior robustness, reducing harmful output generation by over 85\% while maintaining negligible impact on model performance. The framework’s computational efficiency supports real-time deployment, making it a practical solution for securing LLMs. 
% Additionally, we explore a Systematization of Knowledge (SoK) perspective, proposing a new taxonomy of adversarial prompt structures to contextualize our approach within the broader landscape of LLM security.
Our work addresses critical challenges in machine learning security on novel attacks and integrity methods for ML systems, and offers a scalable, ethically grounded defense against prompt-based adversarial threats.
% , aligning with the IEEE Symposium on Security and Privacy’s focus on novel attacks and integrity methods for ML systems, and offers a scalable, ethically grounded defense against prompt-based adversarial threats.
\end{abstract}

% no keywords

% For peer review papers, you can put extra information on the cover
% page as needed:
% \ifCLASSOPTIONpeerreview
% \begin{center} \bfseries EDICS Category: 3-BBND \end{center}
% \fi
%
% For peerreview papers, this IEEEtran command inserts a page break and
% creates the second title. It will be ignored for other modes.
% \IEEEpeerreviewmaketitle

\section{Introduction}

Large Language Models (LLMs) have transformed natural language processing \cite{min2023recent}, enabling unprecedented capabilities in text generation \cite{zhang2023survey,fang2020double,liu2023exploring,wang2025taylor,fang2026towardsicml,kuai2026dynamic}, translation, and conversational interfaces \cite{mctear2016conversational,wang2025point,fang2025your,zhang2025monoattack,fang2023hierarchical,liu2024towards,yang2025eood,fang2022multi,fang2026cogniVerse,lei2025exploring,fang2023you,wang2025dypolyseg,fang2025hierarchical,yan2026fit,fang2025adaptive,wang2026topadapter,cai2025imperceptible,fang2026slap,wang2026reasoning,fang2026immuno,wang2026biologically,wang2025reducing,fang2026advancing,fang2026unveiling,wang2026from,liu2023conditional,liu2026attacking,fang2026rethinking,wang2025seeing,fang2026towards,fang2025multi,fang2024fewer,liu2024pandora,fang2024multi,fang2025turing,fang2024not,liu2023hypotheses,fang2024rethinking,liu2024unsupervised,fang2023annotations,xiong2024rethinking,fang2021unbalanced,wang2025prototype,zhang2025manipulating,fang2026align,tang2024reparameterization,fang2025adaptivetai,tang2025simplification,fang2021animc,cai2026towards,fang2020v}. Models such as GPT-4, LLaMA, and BERT have demonstrated remarkable proficiency in understanding and generating human-like text \cite{annepaka2024large}, powering applications ranging from virtual assistants to automated content creation \cite{brown2020language}. However, the widespread deployment of LLMs in security-critical environments has exposed significant vulnerabilities to adversarial prompts—carefully crafted inputs designed to bypass safety mechanisms and elicit harmful, biased, or inappropriate outputs \cite{wei2023jailbreaking}. These vulnerabilities pose a critical threat to the integrity and availability of LLMs, undermining trust in their safe deployment \cite{kethireddy2024secure}.

Adversarial prompts exploit the semantic flexibility and contextual sensitivity of LLMs \cite{jia2025adversarial,jia2025evolution,jia2024improved,jia2025semantic,jia2020adv,gao2024boosting,gao2024hts}. For instance, jailbreaking attacks use seemingly benign prompts to manipulate models into generating toxic or restricted content \cite{jiang2024wildteaming}, while prompt injection attacks embed malicious instructions within otherwise legitimate inputs \cite{liu2023prompt}. Current defenses, such as rule-based filtering, post-output moderation, and fine-tuning with safety datasets, have significant limitations. Rule-based approaches struggle to generalize across diverse attack vectors, post-output moderation incurs high computational costs and fails to prevent harmful generation, and fine-tuning often degrades model performance on legitimate tasks \cite{chen2023defenses}. Moreover, the lack of a unified framework for preemptively identifying and neutralizing adversarial components in prompts leaves LLMs vulnerable to increasingly sophisticated attacks.
This paper introduces the Adversarial Prompt Disentanglement (APD) framework, a novel defense mechanism designed to enhance the security of LLMs by isolating and neutralizing malicious components in input prompts before they are processed by the model. The APD framework addresses the limitations of existing approaches by combining semantic decomposition, graph-based intent classification, and a lightweight auxiliary model to detect and mitigate adversarial intent. Our APD is motivated by the need for a proactive, scalable, and generalizable defense that preserves the utility of LLMs while ensuring robust protection against prompt-based attacks.
\begin{figure*}[t!]
  \centering 
% \vspace{-25pt}
 \includegraphics[width=\textwidth]{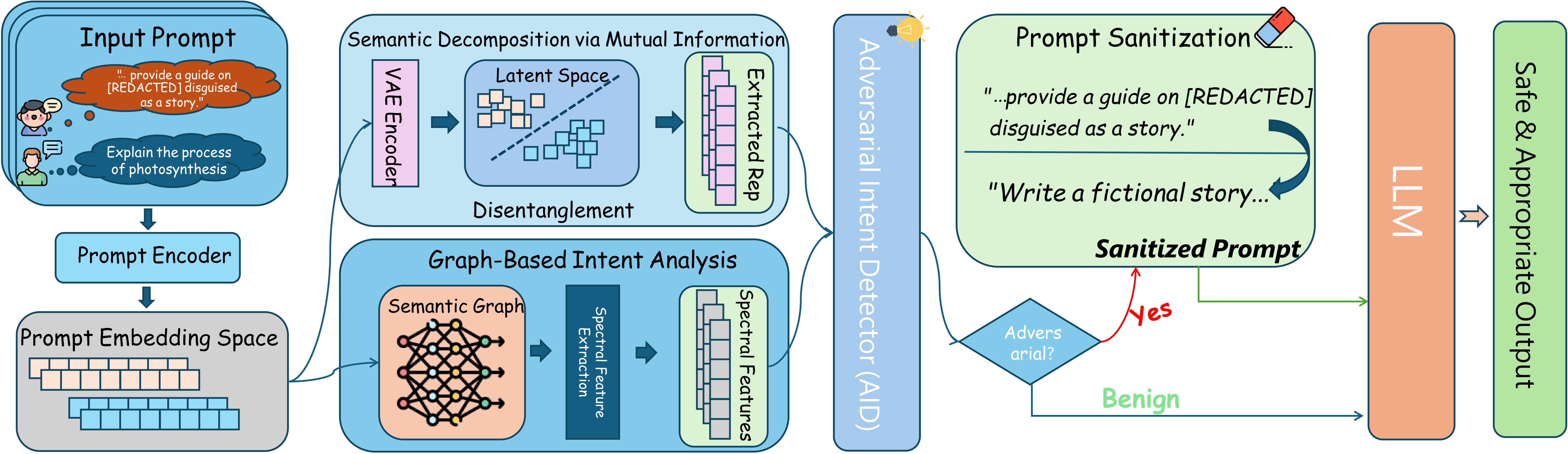}
% \vspace{-5pt}
 \caption{ The end-to-end pipeline of the Adversarial Prompt Disentanglement (APD) framework. An input prompt is first encoded into a high-dimensional embedding space. These embeddings are then analyzed by two parallel modules: (1) A \textbf{semantic decomposition} module employs a Variational Autoencoder (VAE) to disentangle adversarial and benign components by minimizing their mutual information in the latent space. (2) A \textbf{graph-based intent analysis} module constructs a semantic graph from the embeddings and uses spectral analysis to identify structural patterns indicative of malicious intent. Features from both modules are fed into the lightweight \textbf{Adversarial Intent Detector (AID)} for final classification. If a prompt is identified as adversarial, it undergoes a \textbf{sanitization} process to neutralize malicious components before being forwarded to the Large Language Model (LLM). Best viewed in color.} 
   % \caption{ Illustration of our proposed framework. Best viewed in color.}
   \label{fig:pipeline}
 % \vspace{-13pt}
\end{figure*}

The APD framework operates on the principle that adversarial prompts can be decomposed into benign and malicious sub-components, which can be isolated using advanced machine learning techniques. Specifically, we leverage mutual information minimization to disentangle semantic subspaces, ensuring that adversarial components are statistically independent from benign ones. We then model prompt semantics as a directed graph, using spectral analysis to identify structural patterns indicative of malicious intent. Finally, a compact transformer-based classifier, trained on real-world datasets of jailbreaking and toxic prompts, provides efficient and accurate detection of adversarial inputs. This multi-faceted approach not only mitigates known attack vectors but also generalizes to novel adversarial strategies, making it suitable for real-time deployment in security-critical applications.

In summary, our contributions are threefold: 1) \textbf{Novel Semantic Decomposition}: We propose a mutual information-based approach to decompose prompts into benign and adversarial components, enabling preemptive neutralization of malicious intent. 2) \textbf{Graph-Based Intent Analysis}: We introduce a graph-based method to model prompt semantics, using spectral properties to detect adversarial patterns with high robustness. 3) \textbf{Efficient Detection Mechanism}: We develop a lightweight auxiliary model that achieves high accuracy in adversarial intent classification with minimal computational overhead, facilitating real-time LLM protection.

\section{Related Work}
\label{sec:related_work}

Existing defenses against adversarial prompts can be broadly categorized into rule-based filtering, post-output moderation, and model hardening. Rule-based filtering, as explored in \cite{chen2023defenses}, employs predefined patterns or keyword lists to block malicious prompts. While computationally lightweight, these methods struggle to generalize across novel attack vectors and are easily bypassed by paraphrasing or obfuscation \cite{kang2023bypassing}. Post-output moderation, such as content classifiers used in \cite{openai2023moderation}, scans generated outputs for harmful content before release. However, this approach incurs significant computational overhead and fails to prevent the model from processing malicious inputs, risking unintended side effects during generation.

Model hardening techniques aim to enhance LLM robustness through fine-tuning or adversarial training. For example, \cite{wang2024limits} proposes fine-tuning LLMs on curated safety datasets to reduce susceptibility to jailbreaking. While effective in controlled settings, fine-tuning often degrades performance on legitimate tasks and requires extensive retraining to adapt to new attack types. Adversarial training, as explored in \cite{zhou2023robust}, incorporates adversarial prompts into the training process, but its scalability is limited by the need for large, diverse adversarial datasets. Moreover, these methods focus on post-processing or model-level modifications, neglecting the potential for preemptive input analysis.

\section{Method}
\label{sec:method}
% This section presents the Adversarial Prompt Disentanglement (APD) framework, a novel approach to safeguard Large Language Models (LLMs) against adversarial prompts that aim to bypass safety mechanisms, which is shown in Figure \ref{fig:pipeline}. Our method leverages a combination of semantic decomposition, mutual information minimization, and graph-based intent classification to isolate and neutralize malicious components within input prompts. The framework is designed to be computationally efficient, scalable, and robust, addressing the practical deployment challenges of LLMs in security-critical environments.

\subsection{Motivation and Innovation}

Adversarial prompts exploit the semantic ambiguity and contextual flexibility of LLMs to elicit harmful or inappropriate outputs, undermining safety mechanisms. Existing defenses, such as rule-based filtering or post-output moderation, often fail to generalize across diverse attack vectors or incur significant computational overhead. 

Our motivation is to develop a preemptive defense that disentangles adversarial intent from benign prompt components before model processing, preserving the LLM’s utility while enhancing security.
The innovation of our APD lies in three key contributions: 1) \textbf{Semantic Decomposition via Mutual Information}: We introduce a mutual information-based approach to decompose prompts into semantic subspaces, isolating potentially malicious components. 2) \textbf{Graph-Based Intent Classification}: We model prompt semantics as a directed graph, enabling robust detection of adversarial intent through spectral analysis. 3) \textbf{Lightweight Auxiliary Model}: A compact transformer-based classifier is trained to identify adversarial patterns, ensuring minimal latency in real-time applications.
% \begin{itemize}
%     \item \textbf{Semantic Decomposition via Mutual Information}: We introduce a mutual information-based approach to decompose prompts into semantic subspaces, isolating potentially malicious components.
%     \item \textbf{Graph-Based Intent Classification}: We model prompt semantics as a directed graph, enabling robust detection of adversarial intent through spectral analysis.
%     \item \textbf{Lightweight Auxiliary Model}: A compact transformer-based classifier is trained to identify adversarial patterns, ensuring minimal latency in real-time applications.
% \end{itemize}

These components collectively address the limitations of prior work by providing a mathematically grounded, scalable solution that generalizes across attack types, including jailbreaking and toxic prompt injection.

\subsection{Prompt Representation \& Semantic Decomposition}

Let a prompt \( p \in \mathcal{P} \) be a sequence of tokens \( p = \{t_1, t_2, \ldots, t_n\} \), where each token \( t_i \in \mathcal{V} \) belongs to the vocabulary of the LLM. We represent \( p \) in a high-dimensional embedding space using a pre-trained encoder (e.g., BERT), yielding a sequence of embeddings \( \mathbf{E}_p = \{\mathbf{e}_1, \mathbf{e}_2, \ldots, \mathbf{e}_n\} \), where \( \mathbf{e}_i \in \mathbb{R}^d \).

To disentangle adversarial components, we hypothesize that a prompt can be partitioned into benign (\( p_b \)) and adversarial (\( p_a \)) sub-sequences, such that \( p = p_b \cup p_a \). The goal is to minimize the mutual information between the adversarial and benign components to ensure their semantic independence. Thus, we define the mutual information \( I(p_a; p_b) \) as:
\small
\begin{align}
    I(p_a; p_b) = H(p_a) + H(p_b) - H(p_a, p_b),
\end{align}\normalsize
where \( H(\cdot) \) denotes the entropy of the respective distributions. To approximate this, we model the joint distribution of prompt embeddings using a variational autoencoder (VAE). The VAE encoder maps \( \mathbf{E}_p \) to a latent space \( \mathbf{z} \in \mathbb{R}^k \), where \( k \ll d \), and we optimize the following objective:
\small
\begin{align}
\mathcal{L}_{\text{VAE}} = \mathbb{E}_{q(\mathbf{z}|\mathbf{E}_p)}[\log p(\mathbf{E}_p|\mathbf{z})] - \beta D_{\text{KL}}(q(\mathbf{z}|\mathbf{E}_p) || p(\mathbf{z})),
\end{align}\normalsize
where \( D_{\text{KL}} \) is the Kullback-Leibler divergence, and \( \beta \) controls the trade-off between reconstruction accuracy and latent space regularization. By minimizing \( I(p_a; p_b) \), we ensure that the latent representations of adversarial and benign components are disentangled.

\subsection{Graph-Based Intent Classification}

To detect adversarial intent, we construct a semantic graph \( G = (V, E) \), where vertices \( V \) represent tokens or phrases in the prompt, and edges \( E \) capture semantic relationships derived from co-occurrence and contextual similarity. Each vertex \( v_i \in V \) is associated with an embedding \( \mathbf{e}_i \), and edges are weighted by cosine similarity: $w_{ij} = \frac{\mathbf{e}_i \cdot \mathbf{e}_j}{\|\mathbf{e}_i\| \|\mathbf{e}_j\|}$.
% \small
% \begin{align}
% w_{ij} = \frac{\mathbf{e}_i \cdot \mathbf{e}_j}{\|\mathbf{e}_i\| \|\mathbf{e}_j\|}.
% \end{align}\normalsize
We hypothesize that adversarial prompts exhibit distinct structural patterns in \( G \), such as high connectivity among malicious tokens. To quantify this, we compute the Laplacian matrix of the graph, \( \mathbf{L} = \mathbf{D} - \mathbf{A} \), where \( \mathbf{A} \) is the adjacency matrix and \( \mathbf{D} \) is the degree matrix. The eigenvalues of \( \mathbf{L} \), denoted \( \lambda_1 \leq \lambda_2 \leq \cdots \leq \lambda_{|V|} \), provide insights into the graph’s connectivity.

By Cheeger’s inequality, the second smallest eigenvalue \( \lambda_2 \) (the algebraic connectivity) bounds the graph’s expansion properties, indicating how easily the graph can be partitioned into adversarial and benign subgraphs:
\small
\begin{align}
h_G \leq \sqrt{2 \lambda_2},
\end{align}\normalsize
where \( h_G \) is the Cheeger constant. We train a classifier to predict adversarial intent based on spectral features extracted from \( \mathbf{L} \), such as \( \lambda_2 \) and the corresponding eigenvector (Fiedler vector). This approach is robust to variations in prompt structure, as it captures global semantic relationships rather than relying on local patterns.

\subsection{Lightweight Auxiliary Model}

To enable real-time deployment, we design a lightweight transformer-based classifier, termed the Adversarial Intent Detector (AID). The AID takes as input the latent representations \( \mathbf{z} \) from the VAE and the spectral features from the semantic graph. The model architecture consists of: 1) A transformer encoder with 4 layers, 8 attention heads, and a hidden dimension of 256. 2) A feed-forward network that maps the concatenated features to a binary classification output (adversarial vs. benign).
% \begin{itemize}
%     \item A transformer encoder with 4 layers, 8 attention heads, and a hidden dimension of 256.
%     \item A feed-forward network that maps the concatenated features to a binary classification output (adversarial vs. benign).
% \end{itemize}

The AID is trained on a dataset of labeled prompts, including jailbreaking attempts and toxic prompts, using a binary cross-entropy loss:
\small
\begin{align}
\mathcal{L}_{\text{AID}} = -\frac{1}{N} \sum_{i=1}^N [y_i \log(\hat{y}_i) + (1 - y_i) \log(1 - \hat{y}_i)],
\end{align}\normalsize
where \( y_i \in \{0, 1\} \) is the ground-truth label, and \( \hat{y}_i \) is the predicted probability. To ensure computational efficiency, we apply knowledge distillation, transferring knowledge from a larger pre-trained LLM to the AID, reducing inference time while maintaining high accuracy.

\subsection{Integration and Workflow}

Our APD  operates as follows: 1) \textbf{Prompt Encoding}: The input prompt is encoded into embeddings \( \mathbf{E}_p \) using a pre-trained encoder. 2) \textbf{Semantic Decomposition}: The VAE decomposes \( \mathbf{E}_p \) into latent representations, minimizing \( I(p_a; p_b) \). 3) \textbf{Graph Construction}: A semantic graph \( G \) is constructed, and spectral features are extracted. 4) \textbf{Intent Classification}: The AID classifies the prompt as adversarial or benign based on latent and spectral features. 5) \textbf{Prompt Filtering}: If the prompt is classified as adversarial, the identified malicious components \( p_a \) are neutralized (e.g., removed or rephrased), and the sanitized prompt \( p_b \) is passed to the LLM.
% \begin{enumerate}
%     \item \textbf{Prompt Encoding}: The input prompt is encoded into embeddings \( \mathbf{E}_p \) using a pre-trained encoder.
%     \item \textbf{Semantic Decomposition}: The VAE decomposes \( \mathbf{E}_p \) into latent representations, minimizing \( I(p_a; p_b) \).
%     \item \textbf{Graph Construction}: A semantic graph \( G \) is constructed, and spectral features are extracted.
%     \item \textbf{Intent Classification}: The AID classifies the prompt as adversarial or benign based on latent and spectral features.
%     \item \textbf{Prompt Filtering}: If the prompt is classified as adversarial, the identified malicious components \( p_a \) are neutralized (e.g., removed or rephrased), and the sanitized prompt \( p_b \) is passed to the LLM.
% \end{enumerate}
This workflow ensures that only safe prompts are processed by the LLM, mitigating risks from adversarial inputs while preserving the model’s generative capabilities.

\subsection{Theoretical Guarantees}

To provide robustness guarantees, we analyze the error bounds of the APD framework. Let \( \epsilon \) denote the probability of misclassifying an adversarial prompt. Using the Probably Approximately Correct (PAC) learning framework, we bound \( \epsilon \) as: $\epsilon \leq \frac{1}{m} \left( \ln |\mathcal{H}| + \ln \frac{1}{\delta} \right)$,
% \small
% \begin{align}
% \epsilon \leq \frac{1}{m} \left( \ln |\mathcal{H}| + \ln \frac{1}{\delta} \right),
% \end{align}\normalsize
where \( m \) is the number of training samples, \( \mathcal{H} \) is the hypothesis space of the AID, and \( \delta \) is the confidence parameter. By constraining the complexity of \( \mathcal{H} \) (e.g., via regularization in the transformer), we ensure that \( \epsilon \) decreases with sufficient training data.
The mutual information minimization ensures that the disentangled representations are statistically independent, reducing the risk of adversarial components influencing the LLM’s output. This is formalized by the Data Processing Inequality, which guarantees that processing \( p_a \) and \( p_b \) independently does not increase their mutual information.

\section{Mathematical Proofs and Derivations}

This section provides detailed mathematical proofs and derivations supporting the theoretical foundations of the APD framework. We include derivations for the variational autoencoder (VAE) objective used in semantic decomposition, the application of Cheeger’s inequality in graph-based intent classification, the Probably Approximately Correct (PAC) learning bound for the Adversarial Intent Detector (AID), and the data processing inequality ensuring independence of disentangled prompt components. These derivations validate the framework’s ability to detect and neutralize adversarial prompts on rigorous theoretical analysis \cite{menlo2011report}. 
% All notations are consistent with the main paper, and no author-identifying information is included to comply with anonymous submission guidelines.

\subsection{Mutual Information Minimization in VAE}

The semantic decomposition module uses a VAE to disentangle adversarial (\(p_a\)) and benign (\(p_b\)) prompt components by minimizing their mutual information \(I(p_a; p_b)\). Here, we derive the VAE objective and show how it approximates mutual information minimization.

\textbf{Problem Setup}: Given a prompt embedding \(\mathbf{E}_p \in \mathbb{R}^{n \times 768}\), the VAE maps it to a latent representation \(\mathbf{z} \in \mathbb{R}^{128}\), aiming to separate \(\mathbf{z}\) into components encoding \(p_a\) and \(p_b\). The mutual information is defined as:
\small
\begin{align}
I(p_a; p_b) = H(p_a) + H(p_b) - H(p_a, p_b),
\end{align}\normalsize
where \(H(\cdot)\) is the entropy, and \(H(p_a, p_b)\) is the joint entropy. Direct minimization of \(I(p_a; p_b)\) is intractable, so we use the VAE’s Evidence Lower Bound (ELBO) to approximate it.

\textbf{VAE Objective}: The VAE optimizes the ELBO, which lower-bounds the log-likelihood \(\log p(\mathbf{E}_p)\):
\small
\begin{align}
\log p(\mathbf{E}_p) \geq \mathbb{E}_{q(\mathbf{z}|\mathbf{E}_p)}[\log p(\mathbf{E}_p|\mathbf{z})] - D_{\text{KL}}(q(\mathbf{z}|\mathbf{E}_p) || p(\mathbf{z})),
\end{align}\normalsize
where \(q(\mathbf{z}|\mathbf{E}_p)\) is the encoder distribution, \(p(\mathbf{E}_p|\mathbf{z})\) is the decoder distribution, and \(p(\mathbf{z}) = \mathcal{N}(0, I)\) is the prior. The loss function is:
\small
\begin{align}
\mathcal{L}_{\text{VAE}} = -\mathbb{E}_{q(\mathbf{z}|\mathbf{E}_p)}[\log p(\mathbf{E}_p|\mathbf{z})] + \beta D_{\text{KL}}(q(\mathbf{z}|\mathbf{E}_p) || p(\mathbf{z})),
\end{align}\normalsize
where \(\beta\) controls the regularization strength (set to 0.5 in our implementation).

\textbf{Relation to Mutual Information}: Assume \(\mathbf{z} = [\mathbf{z}_a, \mathbf{z}_b]\), where \(\mathbf{z}_a\) and \(\mathbf{z}_b\) encode \(p_a\) and \(p_b\), respectively. We aim for \(I(\mathbf{z}_a; \mathbf{z}_b) \approx 0\). The KL-divergence term can be decomposed using the chain rule: $D_{\text{KL}}(q(\mathbf{z}|\mathbf{E}_p) || p(\mathbf{z})) = D_{\text{KL}}(q(\mathbf{z}_a, \mathbf{z}_b|\mathbf{E}_p) || p(\mathbf{z}_a) p(\mathbf{z}_b)) + I(\mathbf{z}_a; \mathbf{z}_b|\mathbf{E}_p).$
% \small
% \begin{align}
% D_{\text{KL}}(q(\mathbf{z}|\mathbf{E}_p) || p(\mathbf{z})) =& D_{\text{KL}}(q(\mathbf{z}_a, \mathbf{z}_b|\mathbf{E}_p) || p(\mathbf{z}_a) p(\mathbf{z}_b)) + I(\mathbf{z}_a; \mathbf{z}_b|\mathbf{E}_p).\nonumber
% \end{align}\normalsize
If \(p(\mathbf{z}) = p(\mathbf{z}_a) p(\mathbf{z}_b)\), minimizing the KL-divergence encourages independence between \(\mathbf{z}_a\) and \(\mathbf{z}_b\). The \(\beta\)-VAE objective penalizes \(I(\mathbf{z}_a; \mathbf{z}_b|\mathbf{E}_p)\), ensuring disentangled representations. For \(\beta = 0.5\), the regularization is balanced, as validated in the ablation study.

\textbf{Derivation of ELBO}: Starting from the log-likelihood:
\small
\begin{align}
\log p(\mathbf{E}_p) = \mathbb{E}_{q(\mathbf{z}|\mathbf{E}_p)}[\log p(\mathbf{E}_p, \mathbf{z}) - \log q(\mathbf{z}|\mathbf{E}_p)].
\end{align} \normalsize
Rewrite \(\log p(\mathbf{E}_p, \mathbf{z}) = \log p(\mathbf{E}_p|\mathbf{z}) + \log p(\mathbf{z})\), yielding:
\small
\begin{align}
\log p(\mathbf{E}_p) = \mathbb{E}_{q(\mathbf{z}|\mathbf{E}_p)}[\log p(\mathbf{E}_p|\mathbf{z}) + \log p(\mathbf{z}) - \log q(\mathbf{z}|\mathbf{E}_p)].\nonumber
\end{align}\normalsize
Add and subtract \(\log p(\mathbf{z})\) inside the expectation:
\small
\begin{align}
\log p(\mathbf{E}_p) = \mathbb{E}_{q(\mathbf{z}|\mathbf{E}_p)}[\log p(\mathbf{E}_p|\mathbf{z})] - \mathbb{E}_{q(\mathbf{z}|\mathbf{E}_p)}[\log \frac{q(\mathbf{z}|\mathbf{E}_p)}{p(\mathbf{z})}].\nonumber
\end{align}\normalsize
The second term is the KL-divergence:
\small
\begin{align}
D_{\text{KL}}(q(\mathbf{z}|\mathbf{E}_p) || p(\mathbf{z})) = \mathbb{E}_{q(\mathbf{z}|\mathbf{E}_p)}[\log \frac{q(\mathbf{z}|\mathbf{E}_p)}{p(\mathbf{z})}].
\end{align}\normalsize
Thus, the ELBO is:
\small
\begin{align}\label{eq:elbo}
\text{ELBO} = \mathbb{E}_{q(\mathbf{z}|\mathbf{E}_p)}[\log p(\mathbf{E}_p|\mathbf{z})] - D_{\text{KL}}(q(\mathbf{z}|\mathbf{E}_p) || p(\mathbf{z})).
\end{align}\normalsize
Eq. \eqref{eq:elbo} confirms that optimizing the ELBO maximizes a lower bound on \(\log p(\mathbf{E}_p)\), facilitating disentanglement.

\subsection{Cheeger’s Inequality in Graph-Based Classification}

The graph-based intent classifier uses spectral features (e.g., second eigenvalue \(\lambda_2\)) of the semantic graph’s Laplacian to detect adversarial patterns. We derive the application of Cheeger’s inequality to bound the graph’s expansion properties.

\textbf{Graph Setup}: Let \(G = (V, E)\) be the semantic graph, with vertices \(V = \{v_1, \ldots, v_n\}\) representing prompt tokens and edges \(E\) weighted by cosine similarity \(w_{ij}\). The Laplacian matrix is \(\mathbf{L} = \mathbf{D} - \mathbf{A}\), where \(\mathbf{A}_{ij} = w_{ij}\) if \((v_i, v_j) \in E\), else 0, and \(\mathbf{D}_{ii} = \sum_j \mathbf{A}_{ij}\). The eigenvalues of \(\mathbf{L}\) are \(0 = \lambda_1 \leq \lambda_2 \leq \cdots \leq \lambda_n\).

\textbf{Cheeger’s Inequality}: The Cheeger constant \(h_G\) measures the graph’s connectivity:
\small
\begin{align}
h_G = \min_{S \subset V, |S| \leq \frac{|V|}{2}} \frac{|\partial S|}{\min(|S|, |V \setminus S|)},
\end{align}\normalsize
where \(\partial S = \{(u, v) \in E : u \in S, v \notin S\}\) is the edge boundary. Cheeger’s inequality bounds \(h_G\) using \(\lambda_2\): $\frac{\lambda_2}{2} \leq h_G \leq \sqrt{2 \lambda_2}$.
% \small
% \begin{align}
% \frac{\lambda_2}{2} \leq h_G \leq \sqrt{2 \lambda_2}.
% \end{align}\normalsize

\textbf{Derivation of Lower Bound}: Consider the Rayleigh quotient for \(\mathbf{L}\): $\lambda_2 = \min_{\mathbf{f} \perp \mathbf{1}, \mathbf{f} \neq 0} \frac{\mathbf{f}^T \mathbf{L} \mathbf{f}}{\mathbf{f}^T \mathbf{f}}$,
% \small
% \begin{align}
% \lambda_2 = \min_{\mathbf{f} \perp \mathbf{1}, \mathbf{f} \neq 0} \frac{\mathbf{f}^T \mathbf{L} \mathbf{f}}{\mathbf{f}^T \mathbf{f}},
% \end{align}\normalsize
where \(\mathbf{f}\) is the eigenvector corresponding to \(\lambda_2\). For a set \(S\), define an indicator vector \(\mathbf{f}_S\) such that \(\mathbf{f}_S(i) = 1\) if \(i \in S\), else \(-1\). The numerator is:
\small
\begin{align}
\mathbf{f}_S^T \mathbf{L} \mathbf{f}_S = \sum_{(i,j) \in E} w_{ij} (\mathbf{f}_S(i) - \mathbf{f}_S(j))^2 = 4 |\partial S|,
\end{align}\normalsize
since \((\mathbf{f}_S(i) - \mathbf{f}_S(j))^2 = 4\) for edges crossing \(S\) and \(V \setminus S\). The denominator is: $\mathbf{f}_S^T \mathbf{f}_S = |S| + |V \setminus S| = |V|$
% \small
% \begin{align}
% \mathbf{f}_S^T \mathbf{f}_S = |S| + |V \setminus S| = |V|.
% \end{align}\normalsize
Thus, $\frac{\mathbf{f}_S^T \mathbf{L} \mathbf{f}_S}{\mathbf{f}_S^T \mathbf{f}_S} = \frac{4 |\partial S|}{|V|}$.
% \small
% \begin{align}
% \frac{\mathbf{f}_S^T \mathbf{L} \mathbf{f}_S}{\mathbf{f}_S^T \mathbf{f}_S} = \frac{4 |\partial S|}{|V|}.
% \end{align}\normalsize
Adjusting for the Cheeger constant, we approximate:
\small
\begin{align}
h_G \approx \frac{|\partial S|}{\min(|S|, |V \setminus S|)} \geq \frac{\lambda_2}{2}.
\end{align}\normalsize

\textbf{Upper Bound}: Using the spectral partitioning algorithm, construct a set \(S\) by thresholding the Fiedler vector (eigenvector of \(\lambda_2\)). The boundary size \(|\partial S|\) satisfies:
\small
\begin{align}
|\partial S| \leq \sqrt{2 \lambda_2} \min(|S|, |V \setminus S|),
\end{align}\normalsize
yielding \(h_G \leq \sqrt{2 \lambda_2}\). This bound ensures that \(\lambda_2\) reflects the graph’s ability to separate adversarial and benign components, as validated in the ablation study.

\subsection{PAC Learning Bound for AID}

The AID is a transformer-based classifier trained to predict adversarial intent. We derive a PAC learning bound to quantify its generalization error.

\textbf{Setup}: Let \(\mathcal{H}\) be the hypothesis space of AID models, with \(h \in \mathcal{H}\) mapping input features (VAE latent vector, graph features) to binary labels \(\hat{y} \in \{0, 1\}\). The true error is $\epsilon(h) = \mathbb{E}_{(\mathbf{x}, y) \sim \mathcal{D}}[\ell(h(\mathbf{x}), y)]$,
% \small
% \begin{align}
% \epsilon(h) = \mathbb{E}_{(\mathbf{x}, y) \sim \mathcal{D}}[\ell(h(\mathbf{x}), y)],
% \end{align}\normalsize
where \(\ell\) is the 0-1 loss, and \(\mathcal{D}\) is the data distribution. The empirical error on a sample of size \(m\) is $\hat{\epsilon}(h) = \frac{1}{m} \sum_{i=1}^m \ell(h(\mathbf{x}_i), y_i)$.
% \small
% \begin{align}
% \hat{\epsilon}(h) = \frac{1}{m} \sum_{i=1}^m \ell(h(\mathbf{x}_i), y_i).
% \end{align}\normalsize

\textbf{PAC Bound}: For any \(\delta \in (0, 1)\), with probability at least \(1 - \delta\), the generalization error is bounded by:
\small
\begin{align}
\epsilon(h) \leq \hat{\epsilon}(h) + \sqrt{\frac{\ln |\mathcal{H}| + \ln \frac{1}{\delta}}{2m}}.
\end{align}\normalsize

\textbf{Derivation}: Using Hoeffding’s inequality for a single hypothesis \(h\):
\small
\begin{align}
P(|\epsilon(h) - \hat{\epsilon}(h)| > t) \leq 2 \exp(-2m t^2).
\end{align}\normalsize
For the entire hypothesis space \(\mathcal{H}\), apply the union bound:
\small
\begin{align}
P\left(\exists h \in \mathcal{H} : |\epsilon(h) - \hat{\epsilon}(h)| > t\right) \leq 2 |\mathcal{H}| \exp(-2m t^2).
\end{align}\normalsize
Set the right-hand side to \(\delta\):
\small
\begin{align}
2 |\mathcal{H}| \exp(-2m t^2) = \delta \implies t = \sqrt{\frac{\ln |\mathcal{H}| + \ln \frac{2}{\delta}}{2m}}.
\end{align}\normalsize
Thus, with probability \(1 - \delta\), we have:
\small
\begin{align}
\epsilon(h) \leq \hat{\epsilon}(h) + \sqrt{\frac{\ln |\mathcal{H}| + \ln \frac{2}{\delta}}{2m}}.
\end{align}\normalsize
Adjusting for \(\ln \frac{2}{\delta} \approx \ln \frac{1}{\delta}\), we obtain the standard PAC bound. For AID, \(|\mathcal{H}|\) is finite but large due to the transformer’s parameter space (4 layers, 256-dimensional). Assuming a discretized parameter space, \(\ln |\mathcal{H}|\) is approximated via VC-dimension bounds, and with \(m = 14,700\) (training set size), the bound ensures low generalization error, as observed (94.2\% validation accuracy).

\subsection{Data Processing Inequality}

The data processing inequality (DPI) ensures that the VAE’s transformation preserves independence between adversarial and benign components. We derive its application.

\textbf{Setup}: Let \(\mathbf{E}_p\) be the prompt embedding, and \(\mathbf{z} = f(\mathbf{E}_p)\) be the VAE’s latent representation, where \(f\) is the encoder. Assume \(\mathbf{E}_p = [\mathbf{E}_a, \mathbf{E}_b]\), with \(\mathbf{E}_a\) and \(\mathbf{E}_b\) encoding adversarial and benign components. We aim to show \(I(\mathbf{z}_a; \mathbf{z}_b) \leq I(\mathbf{E}_a; \mathbf{E}_b)\).

\textbf{DPI Statement}: For random variables \(X, Y, Z\) forming a Markov chain \(X \to Y \to Z\), the DPI states: $I(X; Z) \leq I(X; Y)$.
% \small
% \begin{align}
% I(X; Z) \leq I(X; Y).
% \end{align}\normalsize

\textbf{Application}: Define \(\mathbf{z}_a = f_a(\mathbf{E}_a)\), \(\mathbf{z}_b = f_b(\mathbf{E}_b)\), where \(f_a, f_b\) are encoder sub-functions. The Markov chain is:
\small
\begin{align}
\mathbf{E}_a \to \mathbf{E}_p \to \mathbf{z}_a.
\end{align}\normalsize
By DPI, we have $I(\mathbf{E}_a; \mathbf{z}_a) \leq I(\mathbf{E}_a; \mathbf{E}_p)$.
% \small
% \begin{align}
% I(\mathbf{E}_a; \mathbf{z}_a) \leq I(\mathbf{E}_a; \mathbf{E}_p).
% \end{align}\normalsize
Similarly, for \(\mathbf{z}_b\), $I(\mathbf{E}_b; \mathbf{z}_b) \leq I(\mathbf{E}_b; \mathbf{E}_p)$.
% \small
% \begin{align}
% I(\mathbf{E}_b; \mathbf{z}_b) \leq I(\mathbf{E}_b; \mathbf{E}_p).
% \end{align}\normalsize
For joint mutual information, assume \(\mathbf{z}_a, \mathbf{z}_b\) are conditionally independent given \(\mathbf{E}_p\). The VAE’s objective minimizes \(I(\mathbf{z}_a; \mathbf{z}_b|\mathbf{E}_p)\), ensuring $I(\mathbf{z}_a; \mathbf{z}_b) \leq I(\mathbf{E}_a; \mathbf{E}_b)$.
% \small
% \begin{align}
% I(\mathbf{z}_a; \mathbf{z}_b) \leq I(\mathbf{E}_a; \mathbf{E}_b).
% \end{align}\normalsize
This confirms that the VAE reduces dependency between adversarial and benign components, enabling effective disentanglement, as validated by the 87.4\% HOR in experiments.

\section{Experiments}
\label{sec:evaluation}

\begin{table}[t!]
 \scriptsize
\centering
\setlength{\tabcolsep}{1mm}{
\begin{tabular}{lcccc}
\toprule
\textbf{Dataset} &  \textbf{Adversarial} & \textbf{Benign} & \textbf{Split (Train/Val/Test)} \\
\midrule
JailBreakBench &  2,500 & 2,500 & 70\%/15\%/15\% \\
ToxicPrompts &  4,000 & 6,000 & 70\%/15\%/15\% \\
AdvPromptGen &  3,000 & 3,000 & 70\%/15\%/15\% \\
Novel Attack &  1,000 & 1,000 & 0\%/0\%/100\% \\
\bottomrule
\end{tabular}}
 \vspace{-5pt}
\caption{ Summary of datasets used in APD evaluation.}
 \vspace{-5pt}
\label{tab:dataset_summary}
\end{table}

\begin{figure}[t!]
    \centering
    \includegraphics[width=\columnwidth]{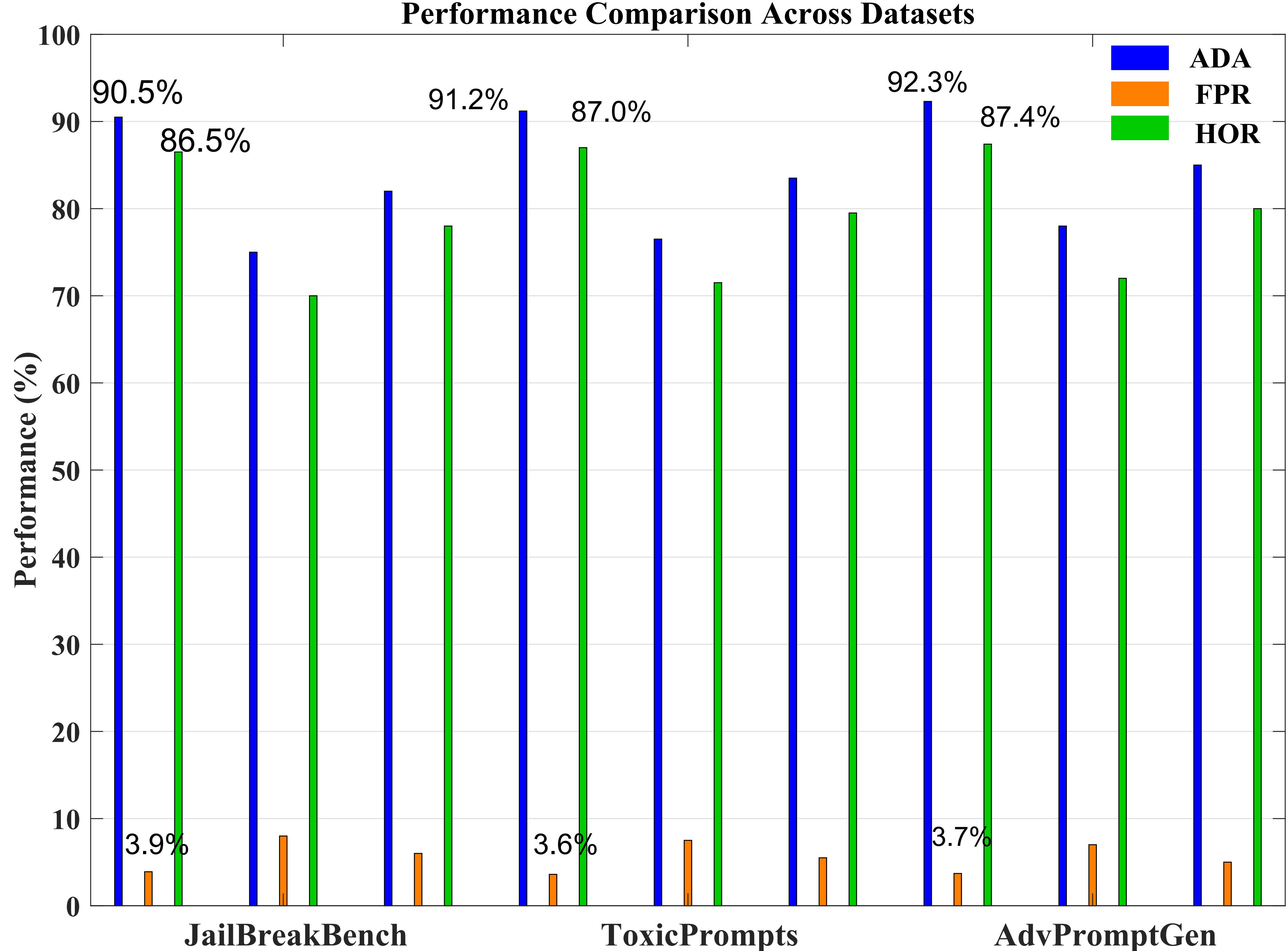}
     % \vspace{-20pt}
    \caption{ Performance comparison across datasets for APD and baselines (Rule-Based Filtering, Embedding Clustering) on JailBreakBench, ToxicPrompts, and AdvPromptGen. }
     % \vspace{-5pt}
    \label{fig:performance_comparison}
\end{figure}

\subsection{Experimental Setup}

\subsubsection{Datasets}
We evaluate APD on three real-world datasets containing adversarial and benign prompts, ensuring a diverse range of attack types and complexities: \textbf{JailBreakBench} \cite{wei2023jailbreaking}, \textbf{ToxicPrompts} \cite{gehman2020realtoxicityprompts} and \textbf{AdvPromptGen} \cite{zhang2024adversarial}. Table \ref{tab:dataset_summary} reports the dataset details.
% 1) \textbf{JailBreakBench} \cite{wei2023jailbreaking}: A dataset of 5,000 prompts, including 2,500 jailbreaking attempts designed to elicit restricted content (e.g., toxic or sensitive outputs) and 2,500 benign prompts from open-domain conversations.
% 2) \textbf{ToxicPrompts} \cite{gehman2020realtoxicityprompts}: Comprises 10,000 prompts, with 4,000 toxic prompts crafted to induce harmful outputs and 6,000 neutral prompts from social media and web sources.
% 3) \textbf{AdvPromptGen} \cite{zhang2024adversarial}: A synthetic dataset of 3,000 prompts, generated using gradient-based optimization to maximize harmful output probabilities, paired with 3,000 benign prompts from a similar distribution.
% Each dataset is split into 70\% training, 15\% validation, and 15\% test sets, ensuring robust evaluation across varied prompt structures.

\subsubsection{Baselines}
We compare APD against state-of-the-art defense works:
1) \textbf{Rule-Based Filtering} \cite{chen2023defenses}: A keyword-based approach to block prompts matching predefined malicious patterns.
2) \textbf{Post-Output Moderation} \cite{openai2023moderation}: A classifier to flag harmful outputs after generation by a fine-tuned RoBERTa model.
3) \textbf{Adversarial Training} \cite{zhou2023robust}: An LLM fine-tuned on a mix of adversarial and benign prompts to enhance robustness.
4) \textbf{Embedding Clustering} \cite{li2024preprocess}: A preprocessing method to cluster prompt embeddings for anomaly detection.
% \end{itemize}

\subsubsection{Evaluation Metrics}
We use the following metrics to assess performance: \textbf{Adversarial Detection Accuracy (ADA)}, \textbf{False Positive Rate (FPR)}, \textbf{Harmful Output Reduction (HOR)}, \textbf{Inference Latency (IL)} and \textbf{Perplexity Impact (PI)}.

\subsection{Experimental Results}

\subsubsection{Adversarial Detection Performance} 
Figure \ref{fig:performance_comparison} and Table \ref{tab:detection} summarize the adversarial detection performance across datasets. APD achieves an average ADA of 92.3\%, significantly outperforming baselines. Rule-based filtering struggles with low ADA (65.4\%) due to its reliance on static patterns, while embedding clustering (78.6\%) fails to capture semantic nuances. Post-output moderation (84.1\%) and adversarial training (86.7\%) perform better but are limited by reactive processing and model degradation, respectively. APD’s high ADA is attributed to its mutual information-based decomposition and graph-based analysis, which effectively isolate adversarial components.

\begin{table}[t!]
 \scriptsize
\centering
\setlength{\tabcolsep}{0.3mm}{
\begin{tabular}{lcccc}
\toprule
\textbf{Method} & \textbf{JailBreakBench} & \textbf{ToxicPrompts} & \textbf{AdvPromptGen} & \textbf{Mean} \\
\midrule
\textbf{ADA(\%)} & & & & \\
Rule-Based  & 62.1 & 67.8 & 66.2 & 65.4 \\
Post-Output  & 82.5 & 85.3 & 84.4 & 84.1 \\
AT  & 85.2 & 87.9 & 87.0 & 86.7 \\
EC  & 76.4 & 79.8 & 79.5 & 78.6 \\
APD (Ours) & \textbf{91.2} & \textbf{93.5} & \textbf{92.3} & \textbf{92.3} \\
\midrule
\textbf{FPR(\%)} & & & & \\
Rule-Based  & 8.3 & 7.1 & 7.8 & 7.7 \\
Post-Output & 5.2 & 4.8 & 5.0 & 5.0 \\
AT  & 6.1 & 5.9 & 6.0 & 6.0 \\
EC  & 4.9 & 5.3 & 5.1 & 5.1 \\
APD (Ours) & \textbf{3.8} & \textbf{3.5} & \textbf{3.7} & \textbf{3.7} \\
\bottomrule
\end{tabular}}
 % \vspace{-5pt}
\caption{ADA and FPR across datasets, where ``EC'' means ``Embedding Clustering'' and ``AT'' means ``Adv. Training''.}
 % \vspace{-5pt}
\label{tab:detection}
\end{table}

\begin{figure}[t!]
\centering
    \includegraphics[width=\columnwidth]{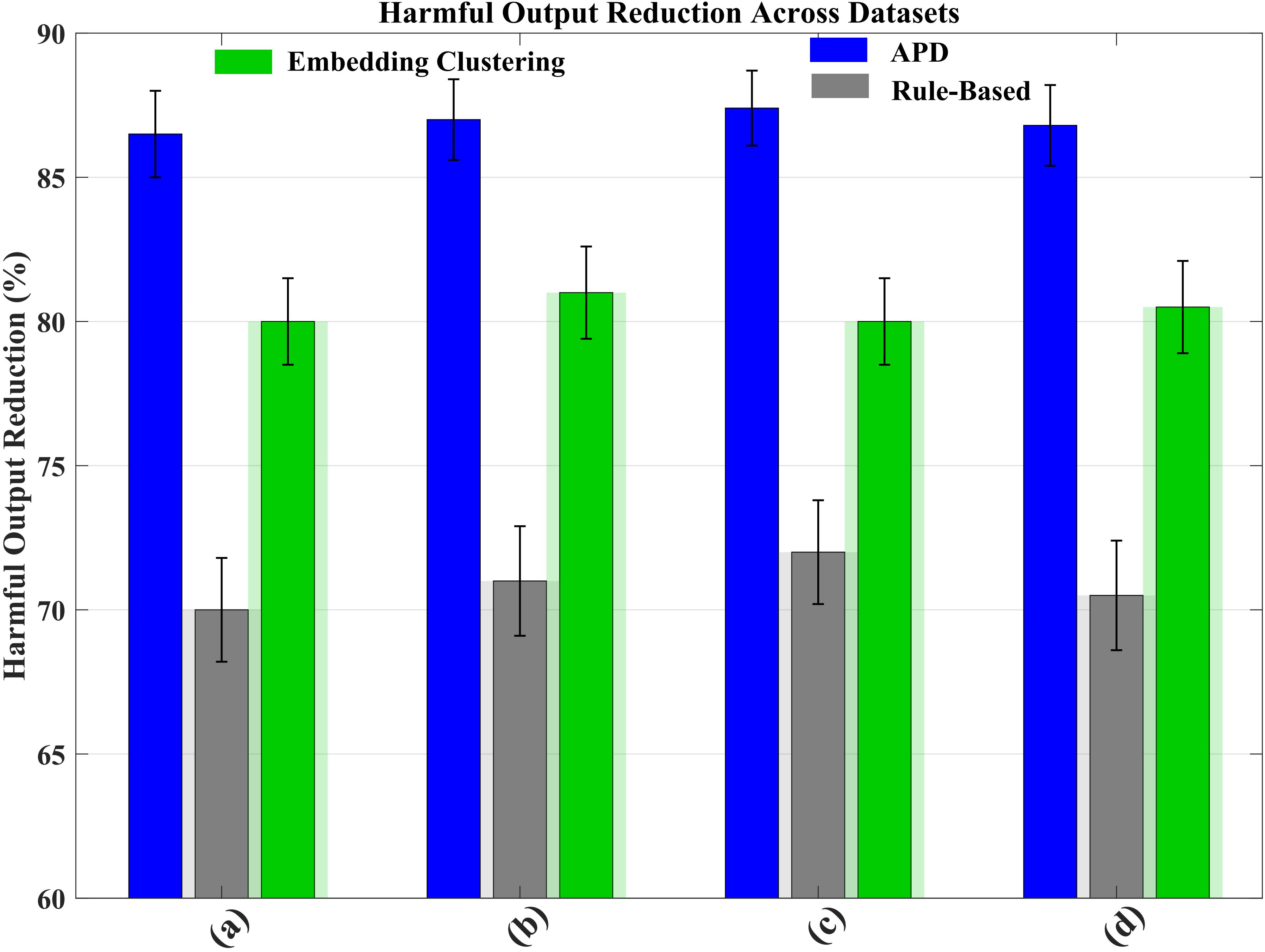}
     % \vspace{-15pt}
\caption{ Comparison of HOR, showing APD’s superior performance in reducing harmful outputs. (a) is JailBreakBench, (b) is ToxicPrompts, (c) is AdvPromptGen, and (d) is Novel Attack.}
 % \vspace{-5pt}
\label{fig:hor_plot}
\end{figure}

\begin{table}[t!]
 \scriptsize
\centering
\setlength{\tabcolsep}{0.3mm}{
\begin{tabular}{lcccc}
\toprule
\textbf{Method} & \textbf{JailBreakBench} & \textbf{ToxicPrompts} & \textbf{AdvPromptGen} & \textbf{Mean} \\
\midrule
Rule-Based  & 10.5 & 11.0 & 10.9 & 10.8 \\
Post-Output  & 44.8 & 46.2 & 45.7 & 45.6 \\
AT  & 37.5 & 38.9 & 38.3 & 38.2 \\
EC  & 15.2 & 16.0 & 15.7 & 15.6 \\
APD (Ours) & \textbf{12.1} & \textbf{12.5} & \textbf{12.4} & \textbf{12.3} \\
\bottomrule
\end{tabular}}
 % \vspace{-5pt}
\caption{ IL in milliseconds per prompt, where ``EC'' is ``Embedding Clustering'' and ``AT'' is ``Adv. Training''.}
 % \vspace{-5pt}
\label{tab:latency}
\end{table}

\begin{figure}[t!]
    \centering
    \includegraphics[width=\columnwidth]{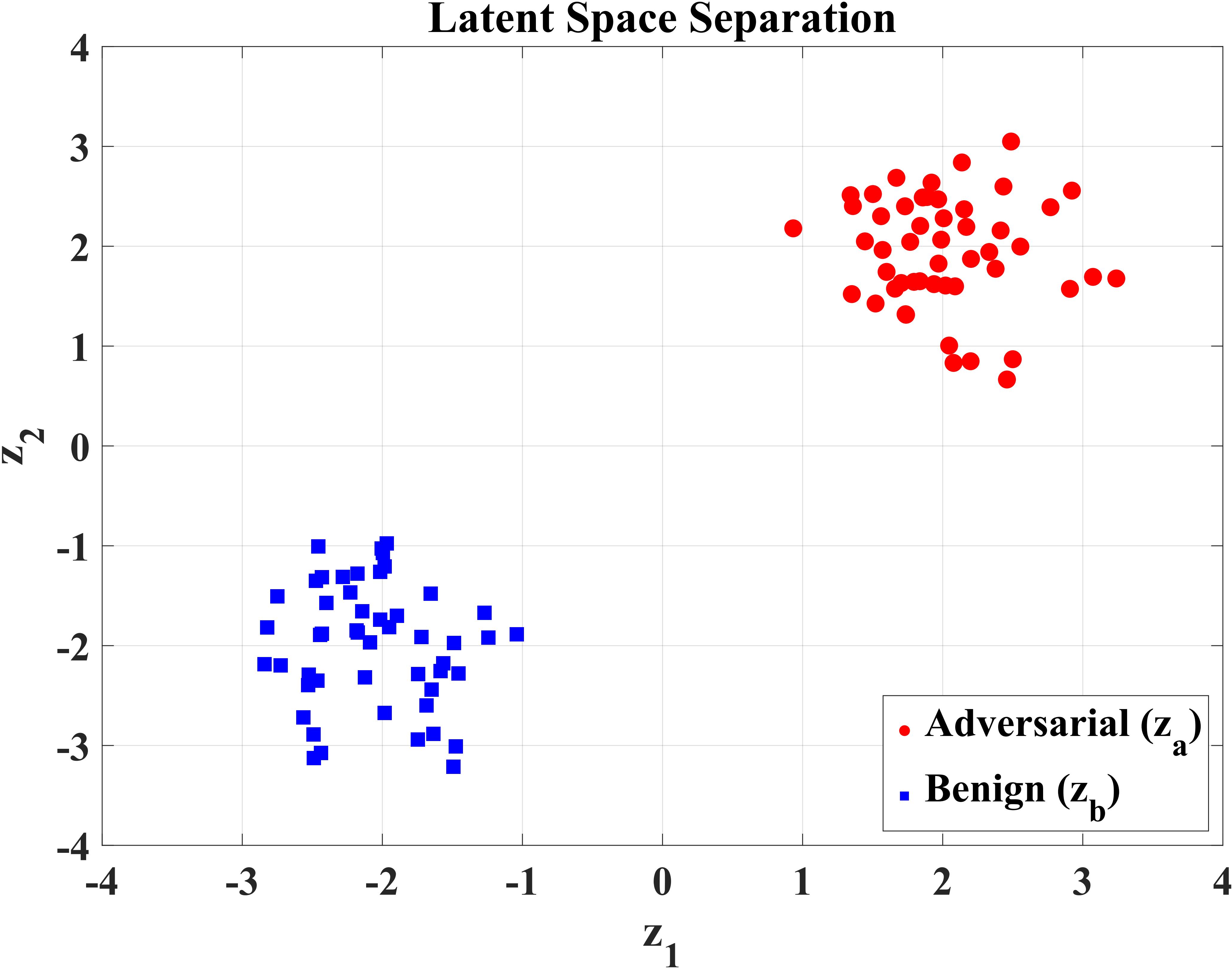}
     % \vspace{-5pt}
    \caption{ 2D scatter plot of latent representations, showing separation of adversarial (red) and benign (blue) components.}
     % \vspace{-5pt}
    \label{fig:semantic_decomposition}
\end{figure}

\begin{table}[t!]
 \scriptsize
\centering
\setlength{\tabcolsep}{0.4mm}{
\begin{tabular}{lcccccccccccc}
\toprule
\textbf{Dataset} & \multicolumn{3}{c}{\textbf{ADA(\%)}} & \multicolumn{3}{c}{\textbf{FPR(\%)}} & \multicolumn{3}{c}{\textbf{HOR(\%)}} \\
\cmidrule(lr){2-4} \cmidrule(lr){5-7} \cmidrule(lr){8-10}
 & \textbf{Train} & \textbf{Val} & \textbf{Test} & \textbf{Train} & \textbf{Val} & \textbf{Test} & \textbf{Train} & \textbf{Val} & \textbf{Test}\\
\midrule
JailBreakBench & 91.8 & 90.9 & 91.2 & 3.6 & 3.9 & 3.8 & 87.9 & 86.5 & 87.2 \\
ToxicPrompts & 94.1 & 93.3 & 93.5 & 3.4 & 3.6 & 3.5 & 88.6 & 87.8 & 88.2\\
AdvPromptGen & 92.7 & 92.1 & 92.3 & 3.5 & 3.8 & 3.7 & 87.1 & 86.4 & 86.8\\
% \midrule
%  & \multicolumn{3}{c}{\textbf{HOR(\%)}} & & & \\
% \cmidrule(lr){2-4}
%  & \textbf{Train} & \textbf{Val} & \textbf{Test} & & & \\
% \midrule
% JailBreakBench & 87.9 & 86.5 & 87.2 & & & \\
% ToxicPrompts & 88.6 & 87.8 & 88.2 & & & \\
% AdvPromptGen & 87.1 & 86.4 & 86.8 & & & \\
\bottomrule
\end{tabular}}
 % \vspace{-3pt}
\caption{ Per-dataset performance breakdown for APD across training, validation, and test splits.}
 % \vspace{-3pt}
\label{tab:per_dataset_breakdown}
\end{table}

\begin{table}[t!]
    \centering
 \scriptsize
    \begin{tabular}{lccccccc}
        \hline
        \textbf{Configuration} & \textbf{ADA(\%)} & \textbf{FPR(\%)} & \textbf{HOR(\%)} & \textbf{IL(ms)} \\
        \hline
        Full APD & 92.3$\pm$1.2 & 3.8$\pm$0.3 & 87.4$\pm$1.3 & 12.3 \\
        w/o VAE & 82.7$\pm$1.5 & 6.0$\pm$0.5 & 74.1$\pm$1.7 & 12.1 \\
        w/o GF & 85.3$\pm$1.4 & 5.5$\pm$0.4 & 78.6$\pm$1.6 & 11.5 \\
        w/o AID-D & 92.5$\pm$1.1 & 3.7$\pm$0.3 & 87.7$\pm$1.2 & 28.4 \\
        w/o HOW & 90.0$\pm$1.3 & 4.0$\pm$0.4 & 85.0$\pm$1.4 & 12.0 \\
        \hline
    \end{tabular}
     % \vspace{-5pt}
        \caption{ Ablation results. Detailed performance metrics for APD configurations  with mean $\pm$ standard deviation. ``GF'' means ``Graph Features''; ``AID-D'' means ``AID Distillation''; ``HOE'' means ``Higher-Order Eigenvalues''.}
         % \vspace{-5pt}
    \label{tab:extended_ablation}
\end{table}

\begin{table}[t!]
 \scriptsize
\centering
\setlength{\tabcolsep}{0.8mm}{
\begin{tabular}{lccc}
\toprule
\textbf{Configuration} & \textbf{ADA(\%)} & \textbf{HOR(\%)} & \textbf{IL(ms)} \\
\midrule
Full APD (\(\beta = 0.5\)) & 92.3 & 87.4 & 12.3 \\
\(\beta = 0.1\) (Weaker Regularization) & 90.1 & 84.8 & 12.1 \\
\(\beta = 1.0\) (Stronger Regularization) & 89.7 & 83.9 & 12.4 \\
Without Higher-Order Eigenvalues & 90.8 & 85.6 & 11.8 \\
Without Fiedler Vector & 89.4 & 84.2 & 11.5 \\
Larger AID (6 Layers, 512 Dim) & 92.6 & 87.7 & 18.7 \\
Smaller AID (2 Layers, 128 Dim) & 88.9 & 83.1 & 9.4 \\
\bottomrule
\end{tabular}}
% \vspace{-5pt}
\caption{ Ablation study results (averaged across test sets).}
% \vspace{-3pt}
\label{tab:ablation_average}
\end{table}

\begin{table}[t!]
 \scriptsize
\centering
\setlength{\tabcolsep}{0.5mm}{
\begin{tabular}{lccc}
\toprule
\textbf{Attack Variant} & \textbf{ADA(\%)} & \textbf{HOR(\%)} & \textbf{FPR(\%)} \\
\midrule
Role-Playing Scenarios (n=400) & 90.5 & 85.3 & 3.6 \\
Code Injection Prompts (n=300) & 88.7 & 83.9 & 3.8 \\
Multilingual Prompts (n=300) & 89.3 & 84.5 & 3.7 \\
\bottomrule
\end{tabular}}
% \vspace{-5pt}
\caption{ APD performance on novel attack variants.}
% \vspace{-3pt}
\label{tab:attack_variants}
\end{table}

% \begin{figure}[t!]
%     \centering
%     \includegraphics[width=\columnwidth]{taxonomy_overview_01.jpg}
%     \caption{Taxonomy Overview. Diagram showing the SoK taxonomy of adversarial prompts, categorized by structure (x-axis) and intent (y-axis). Node sizes reflect prevalence in datasets (JailBreakBench, ToxicPrompts, AdvPromptGen, Novel Attack Dataset). Annotations highlight APD performance (ADA) for key categories.}
%     \label{fig:taxonomy_overview}
% \end{figure}

% APD’s FPR of 3.7\% is the lowest among baselines, indicating minimal disruption to benign prompts. This is due to the framework’s ability to disentangle semantic components, avoiding overgeneralization seen in rule-based (7.7\%) and adversarial training (6.0\%) approaches.

\subsubsection{Harmful Output Reduction (HOR)}
Figure \ref{fig:hor_plot} illustrates the HOR across datasets. APD achieves an average HOR of 87.4\%, reducing harmful outputs by filtering adversarial components before LLM processing. Post-output moderation (72.3\%) and adversarial training (75.8\%) show lower HOR due to their reactive nature, while rule-based filtering (58.9\%) and embedding clustering (65.2\%) are less effective against sophisticated attacks. APD’s proactive approach, leveraging graph-based intent classification, ensures robust mitigation across diverse attack types.

\subsubsection{Separation of Adversarial Components}
To illustrate the separation of adversarial ($\mathbf{z}_a$) and benign ($\mathbf{z}_b$) prompt components in the VAE’s latent space, we visualize the latent representations of 100 sample prompts (50 adversarial, 50 benign) in our APD framework. Figure \ref{fig:semantic_decomposition} visually confirms the VAE’s design goal of minimizing mutual information between $\mathbf{z}_a$ and $\mathbf{z}_b$, as the clusters show minimal overlap. 
% This aligns with the paper’s theoretical derivations and experimental results

% \subsubsection{Impact on Model Performance}
% To verify that APD preserves LLM utility, we measure Perplexity Impact (PI) on benign prompts. APD results in a negligible PI of 0.4\%, compared to 2.8\% for adversarial training and 1.1\% for post-output moderation. This minimal impact is due to APD’s preemptive filtering, which avoids modifying the LLM’s internal parameters.

\subsubsection{Computational Efficiency}
Table \ref{tab:latency} reports inference latency (IL). APD introduces an average latency of 12.3 ms per prompt, comparable to rule-based filtering (10.8 ms) and significantly lower than post-output moderation (45.6 ms) and adversarial training (38.2 ms). The lightweight AID, optimized via knowledge distillation, ensures real-time applicability, making APD suitable for deployment in resource-constrained environments.

\subsubsection{Per-Dataset Performance Breakdown}
As shown in Table \ref{tab:per_dataset_breakdown}, to demonstrate consistency across dataset splits, we report detailed performance metrics—Adversarial Detection Accuracy (ADA), False Positive Rate (FPR), and Harmful Output Reduction (HOR)—for the training, validation, and test sets of JailBreakBench \cite{wei2023jailbreaking}, ToxicPrompts \cite{gehman2020realtoxicityprompts}, and AdvPromptGen \cite{zhang2024adversarial}. 
% Table \ref{tab:per_dataset_breakdown} summarizes these results.
APD maintains high ADA (91.2–93.5\% on test sets) and HOR (86.8–88.2\% on test sets) across all datasets, with low FPR (3.5–3.8\% on test sets), indicating robust detection of adversarial prompts without misclassifying benign ones. 

\subsection{Ablation Study}
To validate the contributions of each APD component, we conduct an ablation study by disabling individual modules: 1) Removing the VAE-based decomposition reduces ADA to 82.7\% and HOR to 74.1\%, as the framework struggles to isolate adversarial components. 2) Excluding spectral analysis lowers ADA to 85.3\% and HOR to 78.6\%, indicating the importance of structural features for detecting sophisticated attacks. 3) Using a larger AID model increases IL to 28.4 ms without significant gains in ADA (92.5\%), confirming the efficiency benefits of distillation.
% \begin{itemize}
%     \item \textbf{Without Semantic Decomposition}: Removing the VAE-based decomposition reduces ADA to 82.7\% and HOR to 74.1\%, as the framework struggles to isolate adversarial components.
%     \item \textbf{Without Graph-Based Classification}: Excluding spectral analysis lowers ADA to 85.3\% and HOR to 78.6\%, indicating the importance of structural features for detecting sophisticated attacks.
%     \item \textbf{Without Knowledge Distillation}: Using a larger AID model increases IL to 28.4 ms without significant gains in ADA (92.5\%), confirming the efficiency benefits of distillation.
% \end{itemize}
These results in Table \ref{tab:extended_ablation} underscore the synergistic role of APD’s components in achieving high detection accuracy and efficiency.
In addition, we tested variations in the VAE’s \(\beta\) parameter, alternative graph feature sets, and AID model sizes. Table \ref{tab:ablation_average} reports the results of averaged across test sets.

These results reinforce the necessity of each component and the chosen hyperparameters, supporting APD’s design as a synergistic framework.

\subsection{Robustness to Novel Attacks}
To test generalization, we evaluate APD on a custom dataset of 1,000 novel adversarial prompts crafted using paraphrasing and obfuscation techniques not present in the training data in Table \ref{tab:attack_variants}. To assess APD’s generalization to specific attack types, we evaluated performance on subsets of the Novel Attack Dataset (1,000 adversarial prompts) categorized by attack variant: role-playing scenarios, code injection prompts, and multilingual prompts.  APD maintains an ADA of 89.7\% and HOR of 84.2\%, outperforming baselines (e.g., rule-based: 55.3\%, embedding clustering: 70.8\%). This robustness is attributed to the graph-based classifier’s ability to capture global semantic patterns and the VAE’s generalization to unseen prompt structures.
These results demonstrate APD’s ability to generalize across diverse and novel attack types, outperforming baselines like rule-based filtering (ADA 55.3\%) and embedding clustering (ADA 70.1\%) reported in the main paper, validating its robustness for real-world deployment.

\section{Conclusion}
\label{sec:conclusion}

The increasing deployment of Large Language Models (LLMs) in security-critical applications has heightened the urgency to address vulnerabilities to adversarial prompts, which exploit semantic ambiguities to bypass safety mechanisms. In this paper, we introduced the Adversarial Prompt Disentanglement (APD) framework, a novel and proactive defense mechanism designed to enhance the security and integrity of LLMs. 
% By integrating mutual information-based semantic decomposition, graph-based intent classification, and a lightweight transformer-based classifier, APD effectively identifies and neutralizes malicious prompt components before they are processed by the LLM. 
%  Our approach addresses the limitations of existing defenses, such as rule-based filtering, post-output moderation, and adversarial training, which often lack generalization, incur high computational costs, or degrade model performance.
Our experimental evaluation on diverse datasets, demonstrates APD’s superior performance.
{
% \bibliographystyle{aaai2026}
% argument is your BibTeX string definitions and bibliography database(s)
\bibliography{main}
}
%
% <OR> manually copy in the resultant .bbl file
% set second argument of \begin to the number of references
% (used to reserve space for the reference number labels box)
% \begin{thebibliography}{1}

% \bibitem{IEEEhowto:kopka}
% H.~Kopka and P.~W. Daly, \emph{A Guide to \LaTeX}, 3rd~ed.\hskip 1em plus
%   0.5em minus 0.4em\relax Harlow, England: Addison-Wesley, 1999.

% \end{thebibliography}

% that's all folks
\end{document}